\documentclass{jpsj2} 
\usepackage{graphicx}
\def\beq{\begin{equation}}
\def\eeq{\end{equation}}
\def\bes{\begin{split}}
\def\ees{\end{split}}
\def\beqa{\begin{eqnarray}}
\def\eeqa{\end{eqnarray}}
\def\ua{\uparrow}
\def\da{\downarrow}

\newcommand{\df}{d}

\newcommand{\etal} {{\it et al.}}
\newcommand{\hc}{{\rm h.c.}}
\newcommand{\hex}{h_{\rm ex}(T)}

\newcommand{\hg}{HgBa$_2$Ca$_4$Cu$_5$O$_y$}
\newcommand{\hl}{H_{\rm L}}
\newcommand{\hr}{H_{\rm R}}

\newcommand{\ic}{I_{\rm c}}
\newcommand{\iu}{{\rm i}}
\newcommand{\kbt}{T}

\newcommand{\kbm}{\mbox{\boldmath $k$}}
\newcommand{\kbms}{\mbox{\boldmath $k$}_{\sigma}}

\newcommand{\kld}{\mbox{\boldmath $k$}_{\rm L\da}}
\newcommand{\kls}{\mbox{\boldmath $k$}_{\rm L\sigma}}
\newcommand{\klu}{\mbox{\boldmath $k$}_{\rm L\ua}}

\newcommand{\kmd}{\mbox{\boldmath $k$}_{\rm FM\da}}
\newcommand{\kms}{\mbox{\boldmath $k$}_{\rm FM\sigma}}
\newcommand{\kmsm}{\mbox{\boldmath $k$}_{\rm FM-\sigma}}
\newcommand{\kmu}{\mbox{\boldmath $k$}_{\rm FM\ua}}

\newcommand{\krd}{\mbox{\boldmath $k$}_{\rm R\da}}
\newcommand{\krs}{\mbox{\boldmath $k$}_{\rm R\sigma}}
\newcommand{\kru}{\mbox{\boldmath $k$}_{\rm R\ua}}
\newcommand{\klr}{\mbox{\boldmath $k$}_{\rm L(R)\sigma}}
\def\lsim{\lower -0.3ex \hbox{$<$} \kern -0.75em \lower 0.7ex 
          \hbox{$\sim$}}
\def\gsim {\raise.35ex\hbox{$>$}\kern-0.75em\lower.5ex\hbox{$\sim$}}

\newcommand{\pst}{$\pi${\it -state}}
\newcommand{\qbm}{\mbox{\boldmath $q$}}
\newcommand{\qqbm}{\mbox{\boldmath $Q$}}
\newcommand{\rbm}{\mbox{\boldmath $r$}}
\newcommand{\rl}{\mbox{{\boldmath $r$}}_L}
\newcommand{\rr}{\mbox{{\boldmath $r$}}_R}

\newcommand{\rmd}{{\rm d}}
\newcommand{\rme}{{\rm e}}

\newcommand{\tauimp}{\tau_{\rm imp}}
\newcommand{\taus}{\tau_{\rm s}(T)}
\newcommand{\tc} {T_{\rm SC}}
\newcommand{\tfm} {T_{\rm FM}}

\newcommand{\vf}{v_{\rm F}}
\newcommand{\wn}{\omega_n}
\newcommand{\xit}{\xi_{\rm T}}

\newcommand{\zst}{{\it 0-state}}
\def\jnl#1#2#3#4{#1 {\bf #2} (#4) #3.}

\def\IJMPB{Int.\ J.\ Mod.\ Phys. B}

\def\JLTP{J.\ Low \ Temp. \ Phys.}
\def\JPSJ{J.\ Phys.\ Soc.\ Jpn.}

\def\PR{Phys.\ Rev.}
\def\PRB{Phys.\ Rev.\ B}
\def\PRL{Phys.\ Rev.\ Lett.}
\def\PTP{Prog.\ Theor.\ Phys.}
\def\RMP{Rev.\ Mod.\ Phys.}
\def\SJETP{Sov.\ Phys.\ JETP}
\title{Role of magnetic scattering in 0-$\pi$ transitions in a superconductor/ferromagnetic metal/superconductor junction}
\author{M. Mori$^{1}$, S. Hikino$^{1}$, S. Takahashi$^{1}$, S. Maekawa$^{1,2}$}
\inst{	
$^{1}$Institute for Materials Research, Tohoku University, Sendai 980-8577, Japan\\
$^{2}$CREST, Japan Science and Technology Agency, Kawaguchi 433-0012, Japan
}

\abst{
We study the influence of magnetic scattering on the Josephson critical current, $\ic$, in a superconductor/ferromagnetic metal/superconductor (SFS) junction by a tunneling Hamiltonian approach. 
An analytical formula of $\ic$ is given in the fourth order perturbation theory as regards the tunneling matrix element. 
The $\ic$ exhibits the damped oscillatory dependence on the thickness of the ferromagnetic metal, $\df$, and shows the transition between ${\it 0}$- and $\pi$-{\it states} with $\df$.
When the superconducting transition temperature is comparable to the ferromagnetic Curie temperature, the period of oscillation is obviously changed by increasing temperature, $T$, due to the magnetic scattering, which induces the $0$-$\pi$ transition with $T$. 
The magnetic scattering provides rich variety of Josephson effect in the SFS junction. 
Our results present an appropriate condition of a superconductor and a ferromagnetic metal to control the ${\it 0}$- and the $\pi$-{\it states}. 
}

\begin{document}
\maketitle
\section{Introduction}
The Josephson effect is characterized by the current through a thin insulating layer without voltage-drop.\cite{josephson}  
It is a macroscopic quantum-mechanical phenomenon to preserve the phase coherence between two superconductors. 
A relation between the current, $I$, and the phase di      fference, $\varphi$, is given by $I=I_{\rm c}\sin\varphi$. 
A finite voltage-drop appears for larger values of $I$ than the Josephson critical current, $\ic$, which decreases monotonically with temperature, $T$.\cite{ambegaokar} 
Similar phenomenon is observed in a weak link through a metallic layer due to the proximity effect.\cite{degennes} 
In a superconductor/normal metal/superconductor (SNS) junction, $\ic$~decreases monotonically with the thickness of metallic layer as well as with $T$.\cite{likhalev} 

There are growing interest in a superconductor/ferromagnetic metal/superconductor (SFS) junction,\cite{golubov,buzdin05,bergeret} in which $\ic$~shows a cusp as a function of  thickness of ferromagnetic layer, $\df$, and/or $T$.\cite{ryazanov_01,kontos,sellier03,sellier04,ryazanov_04,bauer,frolov,bell,oboznov,shelukhin,robinson,born,weides} 
Such a non-monotonous behavior is in marked contrast to $\ic$~in the SNS junction. 
Since the cusp originates from taking an absolute value of the oscillation with alternating sign, the current-phase relation (CPR) is considered to be shifted by $\pi$ in a certain range of $\df$ and/or $T$ in the junction from that in the SNS junction. 
It is called \pst, while the conventional Josephson junction has \zst. 
The \pst~is experimentally observed by measuring the CPR or a spontaneous magnetic flux in a superconducting ring including the SFS junction.\cite{bauer,frolov}
The CPR is obtained by minimizing the free energy, $F$, as, 
$I = 2\rme(\partial F/\partial \varphi) = I_{\rm c}\sin\varphi$. 
Then, the \pst~(\zst) has a negative (positive) coupling constant as $\ic$~$<$ 0 ($\ic$~$>$ 0). 
Such a negative coupling constant was originally found in the Josephson effect with a spin-flip process.\cite{kulik66,shiba,bulaevskii77} 

The \pst~in the SFS junction, that was first predicted by Buzdin \etal\cite{buzdin82} is a novel phenomenon obtained by combining a superconductor (SC) with a ferromagnetic metal (FM), although these orders generally compete with each other. 
Recently, many theoretical studies are made on the \pst~in the SFS junction.\cite{golubov,buzdin05,bergeret,buzdin82,buzdin85,buzdin91,demler,bergeret01,buzdin03,buzdin03jetp,cottet,pistolesi,faure,gusakova,volkov,bergeret06,tanaka97sfs,radovic,cayssol04,cayssol05,yamashita} 
So far, semiclassical approaches like the Eilenberger equation\cite{eilenberger} and the Usadel equation\cite{usadel} are adopted by many authors.\cite{buzdin82,buzdin85,buzdin91,demler,bergeret01,buzdin03,buzdin03jetp,cottet,pistolesi,faure,gusakova,volkov,bergeret06} 
The Bogoliubov-de Gennes equation provides another viewpoint on the \pst~in terms of Andreev bound state.\cite{tanaka97sfs,radovic,cayssol04,cayssol05,yamashita} 
A key mechanism is quite similar to that of Fulde-Ferrell-Larkin-Ovchinnikov (FFLO) state.\cite{fulde,larkin} 
Cooper pairs penetrating in the FM acquire a finite center-of-mass momentum proportional to the magnetic exchange splitting, $\hex$, between the up- and down-spin bands. 
As a result, the pair correlation oscillates in the FM as a function of distance from the interface. 
This oscillation provides the $\pi$-phase shift depending on $\df$. 
For example, if $\df$~is a half of the period of the oscillation, the CPR is shifted by $\pi$ from that of a normal Josephson junction. 
Hence, the transition between the 0- and $\pi$ states occurs by changing $\df$. 
However, this period is not determined only by $\hex$ but also by randomness and spin-dependent scattering in the FM.\cite{oboznov,buzdin85,cottet,pistolesi,faure} 
In particular, the latter process plays an important role in the $T$-induced 0-$\pi$ transition.  
It is noted that the \pst~in the SFS junction remains alive even in a dirty FM, 
whereas the FFLO state is fragile against to a tiny amount of randomness.

In this paper, an analytical formula of $\ic$ in the SFS junction is presented in the fourth order perturbation theory by use of tunneling Hamiltonian\cite{aslamazov,awaka,melin} and Green's function with path integral framework. 
We study the influence of magnetic scattering on the $\df$- and $T$-dependences of $\ic$, in which the transition between the $0$- and $\pi$-states occurs. 
It is found that the magnetic scattering is important to observe the 0-$\pi$ transition, in particular, with $T$.
The tunneling Hamiltonian method makes it possible to treat systematically the magnetic scattering from a microscopic viewpoint. 

As an application of the \pst~in the Josephson junction, several authors have proposed a {\it qubit}, that is a quantum coherent two-level system and is utilized in a process of quantum computing.\cite{ioffe,makhlin,yamashita} 
In general, however, the qubit made with the Josephson junction has a short decoherence time due to several couplings to internal degrees of freedom and/or to the environment. 
The SFS junction has more degrees of freedom like a magnetic scattering than the conventional Josephson junction to break the coherence of Cooper pair. 
Our results provide conditions to control the 0- and $\pi$-states and would show a route to realize the solid-state qubit. 

The rest of this paper is organized as follows.
In \S\ref{method}, we briefly summarize our model Hamiltonian and present a formulation of the $\ic$ in the SFS junction. 
In \S\ref{results}, the $\df$-dependence of $\ic$ is calculated for several conditions. 
A role of magnetic scattering is clarified by showing $T$-dependence of $\ic$ for several cases. 
Summary and discussions are given in \S\ref{summary}. 
Below, $\hbar=1$ and $k_{\rm B}=1$ are used in the equations. 
In order to obtain an actual value, $\hbar=6.6\times 10^{-16}$ eV$\cdot$s and $k_{\rm B}=8.6\times 10^{-5}$ eV$\cdot$K$^{-1}$ are used.

\section{Tunneling Hamiltonian approach for Josephson critical current}\label{method}
\subsection{Model Hamiltonian}
We consider $s$-wave SC's and a FM. 
The mean-field Hamiltonian, $H$, is given by
\beqa
H &=&
	H_{\rm SC}+H_{\rm FM}+H_{\rm T},\label{tot-h}\\
H_{\rm SC}
	&=&
	\hl + \hr,\label{sc-h}\\
H_{\rm L}
	&=&	\sum_{\mbox{\tiny $\kls$}} \xi_{\mbox{\tiny $\kls$}} c_{\mbox{\tiny $\kls$}}^\dag	c_{\mbox{\tiny $\kls$}}\nonumber\\
	&&+ \sum_{\mbox{\tiny $\kls$}}\Delta\rme^{\iu \varphi_{\rm L}}~
			c_{\mbox{\tiny $\klu$}}^\dag c_{\mbox{\tiny -$\kld$}} + \hc,\\
\xi_{\mbox{\tiny $\kms$}}
	&=&
		\displaystyle\frac{1}{2m}\kls^2-\mu,\\
H_{\rm R}
	&=&	({\rm L} \rightarrow {\rm R}),\\
H_{\rm FM}
	&=& H_0 + H_{\rm loc} +H_{\rm imp}+H_{\rm mag},\\
H_0
	&=&	\sum_{\mbox{\tiny $\kms$}}\!
		c_{\mbox{\tiny $\kms$}}^\dag
		\left(\frac{1}{2m}\kms^2-\mu-\sigma \hex
		\right)
		c_{\mbox{\tiny $\kms$}},\label{fm-h}\\
\hex
	&=&
		(J_{\rm H}/2)\langle S^z \rangle,\\
H_{\rm loc}
	&=&	J_{\parallel}\sum_{\langle i,j\rangle}  S^z_i\cdot S^z_j,\label{loc}\\
H_{\rm imp}
 &=& \frac{1}{N} \sum_{\mbox{\tiny $\kms$,$\qbm$}} 
 		u(q) c^{\dag}_{\mbox{\tiny $\kms$+$\qbm$}}c_{\mbox{\tiny $\kms$}},\label{imp-h}\\
H_{\rm mag}
 &=& - \frac{J_{\rm H}}{2\sqrt N}\sum_{\mbox{\tiny $\kms$, $q$}} 
 		\sigma \delta S_q^z
 		c_{\mbox{\tiny $\kms$+$\qbm$}}^\dag c_{\mbox{\tiny $\kms$}},\label{mag-h}\\
\delta S_q^z
	&=&
		S_q^z -\langle S_q^z \rangle,\\
H_{\rm T}
	&=& 
	H_{\rm T}^{\rm L} + H_{\rm T}^{\rm R},\label{tunnel}\\
H_{\rm T}^{\rm L}
	&=&
	\sum_{\mbox{\tiny $\kls$,$\kms$}}
		te^{\iu(\kms-\kls)\rl}
		c_{\mbox{\tiny $\kls$}}^\dag
		c_{\mbox{\tiny $\kms$}}
		+ \hc,\\
H_{\rm T}^{\rm R}
	&=& \sum_{\mbox{\tiny $\krs$,$\kms$}}
		te^{\iu(\kms-\krs)\rr}
		c_{\mbox{\tiny $\krs$}}^\dag
		c_{\mbox{\tiny $\kms$}}
		+ \hc.
\eeqa
In eq. (\ref{sc-h}),  $\hl$ ($\hr$) is the Hamiltonian of left (right) SC, where $\xi_{\mbox{\tiny $\kls$}}$ ($\xi_{\mbox{\tiny $\krs$}}$) is the kinetic energy of electrons with wavenumber $\kls$ ($\krs$) and $\mu$ is the chemical potential. 
The electron effective mass is denoted by $m$. 
The superconducting gap and the phase variable are denoted by $\Delta$ and $\varphi_{\rm L}$ ($\varphi_{\rm R}$), respectively. 
The annihilation and creation operators of electron with wavenumber vector $\kbms$ and spin $\sigma$ are denoted by $c_{\mbox{\tiny $\kbms$}}$ and $c_{\mbox{\tiny $\kbms$}}^{\dag} $, respectively. 
Equation (\ref{fm-h}) describes the kinetic energy of electrons in the FM with the magnetic exchange energy, 
$\hex=(J_{\rm H}/2)\langle S^z \rangle$, 
where $\langle S^z \rangle$ is determined by the self-consistent equation as, 
$
\langle S^z \rangle=(1/2)\tanh[2(\tfm/T) \langle S^z \rangle],
$
and $J_{\rm H}$ is a coupling constant between an electron and a localized moment in the FM with $N$ being the number of atomic sites. 
The Curie temperature, $\tfm$, is determined by the ferromagnetic coupling constant, $J_{\parallel}<0$, in eq. (\ref{loc}).
For the ferromagnetically ordered moments in eq. (\ref{loc}), we adopted a mean-field Hamiltonian given by
$J_{\parallel}\sum_{\langle i,j\rangle} \langle S^z_i \rangle\cdot S^z_j$, 
where $\langle i,j\rangle$ indicates the sum of nearest neighbor sites, and $S^z_i$ denotes the $z$ component of the localized spin at the $i$-th site. 
In eq. (\ref{imp-h}), a non-magnetic impurity potential, $u(q)$, in the FM is averaged in the order of $n_i u^2$, where  $n_i$ is the density of non-magnetic impurity and $u(q)$ is assumed to be independent of $q$. 
We adopt eq.(\ref{mag-h}) to describe a magnetic scattering, $\delta S_q^z\equiv S_q^z-\langle S_q^z\rangle$, in the FM, where $q$ means the wavenumber.
The tunneling Hamiltonian is given by $H_{\rm T}$ in eq.(\ref{tunnel}), where $\rl$ ($\rr$) is the position of the interface between the left (right) SC and the FM. 

The current flows in order to minimize the coupling energy between two SC's, and is given by
\beq
I =
	2\rme \frac{\partial F}{\partial \varphi} \equiv I_c \sin\varphi, 
\eeq
where $F$ is the Free energy, and $\varphi\equiv\varphi_{\rm L}-\varphi_{\rm R}$ is a phase difference between two SC's. 
Hence, it is $F$ as a function of $\varphi$ that we need to calculate. 
Details are summarized in Appendix A.

\subsection{Clean system}
In the clean system, the fourth order term of $F$ as regards $t$ is shown in Fig.~\ref{4th}.
Detailed calculations are shown in Appendix A. 
For $\hex/\mu,~ \omega_n/\mu \ll 1$, the analytical form of $\ic$ is given by,
\beqa
I_c R_0
	&=& 
		\Delta^2 \left( \kbt \sum_{\omega_n} 
        \frac{1}{\omega ^{2}_{n}+\Delta ^{2}}
		\rme^{-2|\omega_n|\df/\vf}
						\right)
		\cos\left(\frac{2\hex}{\vf}\df\right), \label{icln2}
\eeqa
where $R_0^{-1}\equiv(mV N_{\rm F})^2 t^4/2$ is a constant determined by the material and the interfaces. 
The density of states at the Fermi energy is denoted by $N_{\rm F}$, and $V$ is the volume of FM. 
The $\ic$ is plotted in Fig.~\ref{ddepcln} for $\vf$=2.5$\times$10$^{5}$ m/s, $h_0$=0.36 eV, $\Delta$=1.5 meV, $T$=4 K, and $\tfm/\tc$=10. 
$h_0$ is defined as, $h_{\rm ex}\mbox{($T$=0K)}\equiv h_0$. 
The ferromagnetic and the superconducting transition temperatures are denoted by $\tfm$ and $\tc$, respectively. 
When $\df$ increases from $\df$=0, in which the ground state should be the 0-state,  $\df$=$\pi\vf/4\hex$ is the thickness for the first minimum of $\ic$. 
It is found that only the ratio of $\hex$ and $\vf$ provides the period of oscillation in the $\ic$-$\df$ curve.  
%
%
If Ni or Co are chosen as the FM, $\tfm$ is much larger than $\tc$, i.e., $\tfm/\tc\gg 1$, and $\hex$ is almost equal to $h_0$ below $\tc$. 
Therefore, the period of oscillation is estimated as, $\pi\vf/h_0$=1.4 nm, for $\vf$=2.5$\times$10$^{5}$ m/s, $h_0$=0.36 eV. 
This result is consistent with experimental results using clean FM.\cite{shelukhin,robinson,born} 
Although another combination of $\vf$=2.5$\times$10$^{6}$ m/s and $h_0$=3.6 eV also provides the same period as $\pi\vf/h_0$=1.4 nm, this value, i.e., $h_0$=3.6 eV, is considered to be too large for $\hex$. 
Then, we will use $\vf$=2.5$\times$10$^{5}$ m/s, below.

The $\ic$ decays with $T$ and/or $\df$, whose dependences are determined after the summation as regards $\wn$ in eq. (\ref{icln2}). 
It can be done as,
\beqa
&&\kbt \sum_{\omega_n}
        \frac{\rme^{-2|\omega_n|\df/\vf}}{\omega ^{2}_{n}+\Delta ^{2}} \nonumber\\
	&=&
		\frac{1}{2\pi\Delta}
			\int_0^{\infty}\rmd x \sin x 
				\left[
				\frac{\pi T}{\Delta}{\rm cosech}\left(\frac{\pi T x}{\Delta}
				+\frac{\df}{\xit}\right)
			  \right],~~~~\label{icln3}
\eeqa
where $\xit\equiv v_{\rm F}/2\pi T$. See also Appendix B. 
By considering eq.~(\ref{icln3}), the $T$-dependence of eq.~(\ref{icln2}) is shown in Fig.~\ref{tdep} for $d$=5 nm and 100 nm. 
If the thermal length, $\xit$, is much smaller than the thickness, $\df$, i.e., $\xit\ll\df$, eq. (\ref{icln3}) is estimated as, 
\beqa
&&		\frac{1}{2\pi\Delta}
			\int_0^{\infty}\rmd x \sin x 
				\left[
				\frac{\pi T}{\Delta}{\rm cosech}\left(\frac{\pi T x}{\Delta}
				+\frac{\df}{\xit}\right)
			  \right]\nonumber\\
&\simeq & \frac{1}{2\pi\Delta}
			\int_0^{\infty}\rmd x ~\sin x \times
				\left[
				\frac{2\pi T}{\Delta}\exp\left(-\frac{\pi T x}{\Delta}-\frac{\df}{\xit}\right)
				\right],\nonumber\\
&\propto& \exp\left(-\frac{\df}{\xit}\right)~~~\mbox{for}~~~ \frac{\df}{\xit}\gg 1. \label{expo}
\eeqa
This result in eq. (\ref{expo}) is shown by the broken line in Fig. \ref{tdep}, which is obtained by replacing the sum of $\wn$ in eq. (\ref{icln2}) with $\omega_{n=0}=\pi T$ as well. 
In the high temperature region defined by $\df/\xit\gg 1$, the Cooper pairs thermally lose their coherence in the FM and then $\ic$ exponentially decays as $\exp(-\df/\xit)$. 
On the other hand, the solid curve in Fig.~\ref{tdep}, which is the $\ic$-$T$ curve for $\df/\xit\ll1$, deviates from the exponential behavior and looks like the $\ic$-$T$ curve in the SIS junction given by
$
\ic =R_{\rm N}^{-1}\Delta \tanh(\Delta/2T).
$
$R_{\rm N}$ denotes normal resistance of insulating barrier.\cite{ambegaokar} 
This similarity means that the Cooper pairs can go through the FM before losing their coherence, if the FM layer is thinner than the thermal length. 

We can show $\ic$ for $\df/\xit \ll1$ at $T$=0 K as,  
\beqa
&& \lim_{T\rightarrow 0}\kbt\sum_{\omega_n}
        \frac{\rme^{-2|\omega_n|\df/\vf}}{\omega ^{2}_{n}+\Delta ^{2}}
= \frac{1}{2\pi\Delta}
			\int_0^{\infty}\rmd x ~	\frac{\sin x}{x+ \df/\xi_0}\nonumber\\
	&=& \frac{1}{2\pi\Delta}
			\left[ -\cos\left(\frac{\df}{\xi_0}\right){\rm si}\left(\frac{\df}{\xi_0}\right)
			+\sin\left(\frac{\df}{\xi_0}\right){\rm ci}\left(\frac{\df}{\xi_0}\right)\right],\\
	&\propto& \frac{1}{\df/\xi_0}~~\mbox{for}~~\df/\xi_0\gg1,
\eeqa
where $\xi_0\equiv \vf/2\Delta$.
It is noted that $\ic$ in the clean system at $T$=0 K decays with $\df$ in the power law as, $\xi_0/\df$.

\subsection{Dirty FM with magnetic scattering}
In a dirty FM including the magnetic- and non-magnetic scatterings, the fourth order term of $F$ is calculated as shown in Fig. \ref{diffuse}. 
The diffusive motion of electrons in the FM is indicated by $\Gamma$, that is given by a sequential sum of the non-magnetic impurity and magnetic scatterings as shown in Fig. 5. 
The double solid lines represent the Green function in the FM with the self-energy, $\Sigma \left( {\omega _{\rm n} } \right) $,  
in the Born approximation given by
\beqa
\Sigma \left( {\omega _{n} } \right) 
	&=&
		- \frac{\iu}{2}\left( {\frac{1 }{{\tau _{{\rm imp}} }} + \frac{1 }{{\taus }}} \right){\mathop{\rm sgn}} \left( {\omega _{n} } \right),\\
\frac{1}{\tauimp}
	&=&
	2\pi N_{\rm F}n_i u^2,\\
\frac{1}{\taus}
	&=&2\pi N_{\rm F} \left({\frac{J_{\rm H}}{2}} \right)^2\left\langle {\delta S^z \delta S^z } \right\rangle\equiv
		 \frac{1}{\tau_{s0}} \left\langle {\delta S^z \delta S^z } \right\rangle.\label{mag-fluc}
\eeqa
The relaxation times by the non-magnetic and magnetic scatterings are denoted by $\tauimp$ and $\taus$, respectively. 
In this study, the spin-spin correlation function is approximated by that of Ising-type mean-field Hamiltonian as,
\beq
\left\langle {\delta S^Z \delta S^Z } \right\rangle  
	\equiv 
		\frac{1}{4}\left[\frac{T}{T_{{\rm FM}}}\cosh^2\left(\frac{2T_{\rm FM}}{T}\langle S_z\rangle\right) - 1\right]^{-1}.
		\label{spin-spin}
\eeq
By  taking the dirty limit of $\wn \tau$, $\hex \tau$, $\vf |Q| \tau$ $\ll$ 1, one can obtain the following equation as,
\beqa
\Gamma
	&=&
		-2\pi N_{\rm F}\sum\limits_{Q} 
		\frac{{\rm e}^{\iu Q\df}}
		{{D {Q}^2  + 2\left| {\omega _{n} } \right| + 2/\taus  \pm \iu 2\hex }},\\
D 
	&=& 
		\frac{1}{3}v_{\rm F}^{\rm 2} \left( {\frac{1}{{\tau _{{\rm imp}} }} + \frac{1}{{\taus}}} \right)^{ - 1} \label{difcon}
\eeqa
where $D$ is the diffusion constant in the FM. 
It should be noted that $D$ contains $\taus$ as well as $\tauimp$, while $D$ has been so far assumed to be constant by other authors. 
We find that the diffusive motion of electron can be changed by $T$ due to the $T$-dependence of $\taus$. 
We will discuss this point in the next section. 

Finally, the $\ic$ in the dirty FM is given by 
\beqa
I_c R_{\rm D}
	&=& \Delta^2 \kbt \sum_{\wn>0}
	    \frac{1}{\omega ^{2}_{n}+\Delta ^{2}}
		\exp\left(-\frac{\df}{\xi_+}\right)
		\cos\left(\frac{\df}{\xi_-}\right),\label{ic}\\
\xi_-
	&\equiv& \!\!\!\!\left[\frac{D}
	{\sqrt{(\omega_n+1/\taus)^2+\hex^2}-(\omega_n+1/\taus)}\right]^{1/2},\label{period} \\
\xi_+
	&\equiv& \!\!\!\!\left[\frac{D}
	{\sqrt{(\omega_n+1/\taus)^2+\hex^2}+(\omega_n+1/\taus)}\right]^{1/2},~~~~
	\label{dump}
\eeqa
where $R_{\rm D}^{-1}\equiv t^4N_{\rm F}^3\pi^2V/2Dd$ is assumed to be a constant determined by the material and the interfaces.
The summation in eq. (\ref{ic}) is numerically carried out, and results are shown in the next section. 

\section{$\df$- and $T$-dependences of 0- and $\pi$-states}\label{results}
We have shown in eq. (\ref{icln2}) that the $\ic$ in the clean system oscillates as a function of $\df$.  
The period is determined by $2\hex/\vf$, and the decay length is given by $\xit$ for $\df/\xit\gg1$. 
On the other hand, in the dirty FM with magnetic scattering, the period, $\xi_-$, and the decay length, $\xi_+$, depend on $\hex$, $D$, and $\taus$, as shown in eqs. (\ref{period}) and (\ref{dump}). 
The $|\ic|$-$d$ curves are plotted in Fig.~\ref{ddep} for 
(a) $T_{\rm FM}/T_{\rm SC}$=10, $h_{0}$=30 meV, $\tauimp$=2$\times10^{-14}$ s, $\tau_{s0}$=$10^{-12}$ s, 
(b) $T_{\rm FM}/T_{\rm SC}$=1,  $h_{0}$=30 meV, $\tauimp$=2$\times10^{-14}$ s, $\tau_{s0}$=$10^{-13}$ s, 
and (c) $T_{\rm FM}/T_{\rm SC}$=1,  $h_{0}$=100 meV, $\tauimp$=2$\times10^{-13}$ s, $\tau_{s0}$=$10^{-13}$ s. 
The other parameters are set as, $\Delta$=1.5 meV, $\tc$=10 K, and  $\vf$=2.5$\times$10$^5$ m/s. 
Although the \pst~is characterized by a negative value of $\ic$, 
$|\ic|$ is useful to compare our results with experimental ones. 

In all cases, $|\ic|$ shows the damped oscillatory behavior as a function of $\df$, and the 0-$\pi$ transition occurs. 
For $h_{0}$=30 meV at $T/\tc$=0.2 in Figs. \ref{ddep} (a) and (b), the first minimum appears around $d$=6.3 nm, which can be considered the case of ferromagnetic metal-alloy, PdNi.\cite{kontos}
If one assumes a larger value of $h_{0}$ as 100 meV, the period of oscillation becomes small. 
However, by taking a smaller value of $\tauimp$ as 2$\times10^{-13}$ s, the first minimum appears around 7.5 nm as shown in Fig. \ref{ddep} (c), since $\xi_-$ is determined by both $\hex$ and $D$ including $\tauimp$. 

For $\tfm/\tc$=10 in Fig. \ref{ddep} (a), the period of oscillation does not change with $T$. 
On the other hand, in the case of $\tfm/\tc$=1, the period is strongly elongated or shortened with $T$ as shown in Figs. \ref{ddep} (b) and (c). 
The difference between the two cases, i.e., $\tfm/\tc$=10 and 1, originates from the $T$-dependence of $\taus$ and $\hex$ in eq.(\ref{period}). 
For $\tfm/\tc$=1, $\taus^{-1}$ ($\hex$) exponentially increases (decreases) with $T$ as shown in Fig. \ref{taus}, while these are almost constant below $\tc$ for $\tfm/\tc$=10. 
Such $T$-dependences are attributed to the magnetic scattering. 
Several authors have pointed out that the magnetic relaxation time elongates the period of $\ic$ and changes the ratio of the decay length and the period.\cite{oboznov,buzdin85,cottet,pistolesi,faure}
Our result includes their one, which corresponds to the case of Fig. \ref{ddep} (b). 
Some experimental studies have reported that the cusp in the $|\ic|$-$\df$ curve is shifted toward a larger value of $d$ with increasing $T$.\cite{ryazanov_01,sellier03,sellier04,ryazanov_04,bauer,frolov,oboznov} 
Figure \ref{ddep} (b), in which $\tfm/\tc$=1 and $\tauimp/\tau_{s0}$=0.1, is consistent with these experimental results. 
On the other hand, in a certain case like $\tauimp/\tau_{s0}$=1, the period of oscillation decreases with $T$, and the minimum of $|\ic|$ is shifted toward zero as shown in Fig. \ref{ddep} (c). 
Although this case is not yet observed in an experiment, it would be realized by controlling the randomness and the magnetism. 

What does make this difference between Figs. \ref{ddep} (b) and (c)? 
Since $D$ is composed of $\tauimp$ and $\taus$ in eq. (\ref{difcon}), the $T$-dependence of $D$ changes depending on the ratio, $\tauimp/\taus$. 
For example, if $1/\taus\ll1/\tauimp$, $D$ can be approximated as a constant, while $D$ exponentially decreases with $T$ for $1/\taus\gg1/\tauimp$. 
This fact means that the ratio of non-magnetic- and magnetic relaxation times, $\tauimp/\taus$, is also an important parameter to control the $T$-dependence of $|\ic|$. 
As we have shown by the perturbative calculation that $\taus$ is included in the denominator of $\xi_-$ and in $D$ in eq. (\ref{period}), 
the period of oscillation, $\xi_-$, shows various behaviors with $T$ and $\tauimp/\taus$. 
In Fig. \ref{xi}, $\xi_-$ is plotted for (a) $h_{0}$=30 meV, $\tauimp$=2$\times10^{-14}$ s, and (b) $h_{0}$=100 meV, $\tauimp$=2$\times10^{-13}$ s. 
Other parameters are set to $\vf$=2.5$\times10^{5}$ m/s, $T_{\rm FM}/T_{\rm SC}$ =1, and $\tau_{s0}$=10$^{-13}$ s. 
In Fig. \ref{xi} (a), which provides Fig. \ref{ddep} (b), $\xi_-$ increases with $T$, while it starts to decrease for $\tauimp>$ 0.5$\times$10$^{-13}$ s in Fig. \ref{xi} (b). 

By setting $\df$ near the thickness of $\ic\sim$ 0, the 0-$\pi$ transition occurs by changing $T$ for $\tfm/\tc$=1, while it is not found for $\tfm/\tc$=10.  
In Fig.~\ref{tdrt}, the $T$-dependences of $|\ic|$ are shown for (a) $\df$=6.5 nm and (b) $\df$=7.5 nm. 
In both cases, the ground state is the \pst.
The solid- and broken lines show $|\ic|$ for $\tfm$/$\tc$=1 and for $\tfm$/$\tc$=10, respectively. 
In the latter case, $|\ic|$ monotonously decreases with $T$ as found in the SIS junction, since the period of oscillation, $\xi_-$, does not change with $T$. 
On  the  other  hand, when $\tfm$ is comparable to $\tc$, $|\ic|$ shows the cusp in the $|\ic|$-$\df$ curve, since $\xi_-$ is strongly elongated or in some cases shorten with $T$ as shown in Figs. \ref{xi} (b) and (c). 
As a result, $\ic$ changes its sign from negative to positive by increasing $T$ and the 0-$\pi$ transition occurs. 
The $T$-induced 0-$\pi$ transition is observed in some experiments,\cite{ryazanov_01,sellier03,sellier04,bauer,oboznov,frolov} which can be attributed to the magnetic scattering leading to the $T$-dependence of $\xi_-$ with $\tfm\sim\tc$. 

We notice that $|\ic|$ in higher $T$ is enhanced compared to that in lower $T$ in Fig.~\ref{tdrt} (a), and vice versa in Fig.\ref{tdrt} (b). 
If we remembered the $|\ic|$-$T$ curve in the SIS- and/or the SNS junctions, Fig.~\ref{tdrt} (a) would seem to be strange. 
However, the thickness in Fig.~\ref{tdrt} (a) is closer to the cusp in the $|\ic|$-$\df$ curve than that in Fig.~\ref{tdrt} (b). 
Thus, due to the large shift of cusp with $T$ in the $|\ic|$-$\df$ curve, Fig.~\ref{tdrt} (a) also becomes possible.  
Actually, both behaviors shown in Figs.~\ref{tdrt} (a) and (b) are observed in the case of Cu$_{52}$Ni$_{48}$ with $\tfm<$ 20 K and Nb with $\tc=$ 9.23 K by Sellier et al.\cite{sellier04} 

The ratios of the period and the decay length with $\omega_{n=0}=\pi T$ is given by 
\beqa
\left(\frac{\xi_-}{\xi_+}\right)^2
	&=&
	1+2\left(\frac{\pi T+\taus^{-1}}{\hex}\right)
	\sqrt{1+\left(\frac{\pi T+\taus^{-1}}{\hex}\right)^2}\nonumber\\
	&+&2\left(\frac{\pi T+\taus^{-1}}{\hex}\right)^2, 
\eeqa
which is plotted in Fig.~\ref{ratio}; (a) by the solid line for 
$T_{\rm FM}/T_{\rm SC}$=10, $h_{0}$=30 meV, $\tauimp$=2$\times10^{-14}$ s, $\tau_{s0}$=$10^{-12}$ s, (b) by the broken line for
$T_{\rm FM}/T_{\rm SC}$=1,  $h_{0}$=30 meV, $\tauimp$=2$\times10^{-14}$ s, $\tau_{s0}$=$10^{-13}$ s, 
and (c) by the dotted broken line for $T_{\rm FM}/T_{\rm SC}$=1,  $h_{0}$=100  meV, $\tauimp$=2$\times10^{-13}$ s, $\tau_{s0}$=$10^{-13}$ s. 
In all cases, $\xi_-/\xi_+$ linearly increases with $T$ around $T/\tc\sim 0$, since $\hex$ and $\taus$ do not change with $T$ and only $\omega_{n=0}=\pi T$ contributes to the $T$-dependence of $\xi_-/\xi_+$.  
For $T_{\rm FM}/T_{\rm SC}$=10, $\xi_-/\xi_+$ linearly increases upto $\tc$ due to $\pi T>1/\taus$. 
On the other hand, in the case of $T_{\rm FM}/T_{\rm SC}$=1, $\xi_-/\xi_+$ are exponentially enhanced below $\tc$ due to $T$-dependence of $\hex$ and $\taus$ as shown in Fig.~\ref{taus}. 
Some authors have claimed that $\xi_-/\xi_+$ is approximately equal to 1 for $\hex\gg T$,\cite{bergeret01,faure,gusakova,volkov} and some experimental studies have reported that $\xi_-/\xi_+\gsim 1$ in the Josephson junction with dirty FM.\cite{bell,oboznov,kontos}
Our results for $\hex\gg T$, $1/\taus$ also show $\xi_-/\xi_+\gsim 1$, which is consistent with those theoretical and experimental results.  
Moreover, we have shown that $\xi_-/\xi_+$ can be enhanced by $T$ due to the magnetic scatterings, 
which provides new possibilities and rich variety for the Josephson effect in the SFS junction as shown in this paper. 

\section{Summary and discussions}\label{summary}
In this paper, we have studied the Josephson critical current, $\ic$, in the superconductor/ferromagnetic metal/superconductor (SFS) junction by a tunneling Hamiltonian method. 
The analytical formula of $\ic$ is presented by the Green's function with path integral framework in the fourth order perturbation theory as regards the tunneling matrix element. 
The influences of magnetic scattering on the $\ic$ are discussed by calculating $\ic$ as a function of thickness of ferromagnetic metal (FM), $\df$, and temperature, $T$. 
The $\ic$ exhibits the damped oscillatory dependence on the thickness, and shows a transition between ${\it 0}$- and $\pi$-{\it states}. 
The oscillation period in the clean system is determined by the magnetic exchange splitting between the up- and down-spin bands in the FM, $\hex$. 
In the dirty FM with magnetic scattering, the period depends on $\hex$, diffusion  constant, and magnetic relaxation time, $\taus$. 
In particular, we found that $\taus$ plays an important role in the temperature-induced 0-$\pi$ transition. 
When the superconducting transition temperature is comparable to the Curie temperature, 
it is found that the period of this oscillation is obviously changed by increasing $T$ due to the magnetic scattering. 

In our approach, the interfaces between FM and superconductors (SC's) are tunnel-like and their order parameters sharply drop at the interface. 
On the other hand, the spatial variation of order parameters on the SC side near the interfaces becomes important for low tunnel-barrier or junction with a metallic interface. 
Since our method does not need any special boundary condition at an interface, 
such a variation can be included by taking higher order terms about the tunneling matrix element and self-consistent calculations. 
However, we will leave it in a future issue. 

Recent progress in the growth techniques for thin films provides various types of magnetic Josephson junctions like a SC/anti-ferromagnet(AFM)/SC junction\cite{bosovic}, and  a SC/half-metallic ferromagnet/SC junction.\cite{pena,keizer} 
Moreover, some of multilayered high-$T_{\rm c}$ cuprates like \hg~inherently contain the magnetic Josephson junction, in which AFM separates two SC's.\cite{bosovic,tokiwa,kotegawa04,mori05,mukuda,b_andersen} 
Related to the high-$T_{\rm c}$ cuprates, the Josephson junction of $d$-wave SC has been  studied by some authors.\cite{tanaka96,tanaka97,kashiwaya00}
Researches on these magnetic Josephson junctions will open up a new possibility of interplay between the Josephson current and magnetism. 
Our method is applicable to such various types of magnetic Josephson junctions.\cite{asano,saburo}

Finally, although the spin scattering is partly included in this study, the spin wave excitation is not considered. 
Such massless excitations will generate additional dissipation in the Josephson current, which is observed as a tiny voltage drop below $\ic$ in the SC/normal metal/SC (SNS) junction.\cite{awaka,ambegaokar69,takayanagi} 
The dissipation in SFS junctions is an interesting research topic in order to realize a solid state qubit. 

\begin{acknowledgments}
 This work was supported by a Grand-in-Aid for Scientific Research on Priority Areas and the NAREGI Nanoscience Project from MEXT and CREST. 
One of authors (M. M.) acknowledges support by 21st Century COE program, Tohoku University and by MEXT.
\end{acknowledgments}

\appendix{
\section{Perturbative calculation of $\ic$ in path integral framework}
\subsection{Basic formula}
In the path integral framework, the partition function is given by 
\beq
 Z = {\rm Tr[e}^{{\rm  - }\beta H} ] 
   = {\rm e}^{{\rm  - }\beta F}
 	 = \int {{\cal D}\Psi ^* {\cal D}\Psi {\rm e}^{- S} }, 
\eeq
where
\beqa
&&S	=
		\sum_{\mbox{\tiny \boldmath $k$}_\sigma,\mbox{\tiny \boldmath $k$}_\sigma',\iu\wn}
		\Psi ^*_{\mbox{\tiny \boldmath $k$}_\sigma}
		\left[ - G_0^{-1}+T \right] 
		\Psi_{\mbox{\tiny \boldmath $k$}_\sigma'},\\
&&\Psi_{\mbox{\tiny \boldmath $k$}_\sigma}
	\equiv 
		\left( 
		\begin{array}{l}
 			\psi _{{\rm L} \uparrow } (\klu,\iu\wn)\\
 			\psi _{{\rm L} \downarrow }^* (-\kld,-\iu\wn)\\
 			\psi _{{\rm FM} \uparrow } (\kmu,\iu\wn)\\
 			\psi _{{\rm FM} \downarrow }^* (-\kmd,-\iu\wn)\\
 			\psi _{{\rm R} \uparrow } (\kru,\iu\wn)\\
 			\psi _{{\rm R} \downarrow }^* (-\krd,-\iu\wn)
 		\end{array}
	  \right),\label{z}\\
&&G_0^{ - 1}  
	= \left( {\begin{array}{*{20}c}
   {G_{\rm L\sigma}^{ - 1} } & 0 & 0  \\
   {\rm 0} & {G_{{\rm FM\sigma}}^{ - 1} } & 0  \\
   {\rm 0} & 0 & {G_{\rm R\sigma}^{ - 1} }  \\
\end{array}} \right).
\eeqa
Each element of $G_0^{-1}$ is the Green's function in the SC, 
\beqa
G_{{\rm L(R)}\sigma}^{ - 1}  
	\!\!\!&=&\!\!\! 
	- \iu\omega_n \sigma_0 - \left( { - \frac{1}{2m}\klr^2  - \mu }\right )\sigma _3\nonumber\\
	&&\!\!\! - \Delta _{{\rm L(R)}} {\rm e}^{\iu\theta _{{\rm L(R)}} \sigma _3 } \sigma _1, 
\eeqa
and that in the FM, 
\beqa
G_{{\rm FM}\sigma}^{ - 1}  
	\!\!\!&=&\!\!\! 
	-\iu\omega_n \sigma_0 -\left( { - \frac{1}{2m}\kms^{\rm 2}  - \mu - \sigma \hex }\right)\sigma_3,~~~~
\eeqa
where $\sigma_a$ ($a$=0, 1, 2, 3) is the Pauli matrix. 
In this framework, the tunneling Hamiltonian is given in the following matrix form as, 
\beqa
T &=& \left( {\begin{array}{*{20}c}
   0 & T_{\rm 12} & 0  \\
   T_{\rm 21} & 0 & T_{\rm 23}\\
   0 & T_{\rm 32} & 0  \\
\end{array}} \right),\\
T_{\rm 12} 
	&=& -t\exp[\iu(\kms-\kls)\rl] \sigma_3=T^*_{\rm 21},\\
T_{\rm 23} 
	&=& -t\exp[\iu(\krs-\kms)\rr] \sigma_3=T^*_{\rm 32}.
\eeqa
The phase factor in the tunneling matrix element originates from the position vector of the interface, since a flat interface between SC and FM is written as,
$
t\delta(\rbm-\rl).
$ 
After integrating $\Psi$ out in eq. (\ref{z}), the free energy is given by
\beq %
F =
		-{\rm Tr}\ln [ - G_0^{-1}+T].\label{ffull}
\eeq
In the fourth order perturbation theory about $t$, $F$ is approximated to be, 
\beqa
F &\simeq&
		{\rm Trln[} G_0^{-1} ]\nonumber\\
	&+&
		{\frac{1}{2}\rm Tr[}G_0 TG_0 T] + \frac{1}{4}{\rm Tr[}G_0 TG_0 TG_0 TG_0 T].\label{f4th}~~~~~~
\eeqa
The first and second terms do not contribute to the Josephson coupling energy, whose leading term is the third one in eq. (\ref{f4th}). 

\subsection{Clean system}
In the clean system, the forth order term is given by
\beqa 
\begin{split}
F_{{\rm eff}} 
	=&2t^4 T{\rm Re}[
	\sum_{\{\mbox{\tiny \boldmath $k$}\},\iu\omega_n,\sigma} {{\rm e}^{{\rm  - }\iu(\kms- \kmsm) \cdot (\rr  - \rl )} }\\
	&
	\times
	f_{\rm R\sigma} (\krs ,\iu\omega_n )
	g^{(0)}_{{\rm FM}\sigma} (\kms,\iu\omega_n )\\
	&
	\times
	g^{(0)}_{{\rm FM}-\sigma} ( - \kmsm  , - \iu\omega_n)
	f_{\rm L\sigma}^{\rm *} (\kls ,\iu\omega_n )],\label{feff}
\end{split}
\eeqa
where each Green's function is given by
\beqa
&&f_{{\rm L(R)}\sigma} (\klr ,\iu\omega_n ) 
= \frac{\Delta _{\rm L(R)}\rme^{\iu \varphi_{\rm L(R)}}}
         {\omega_n^{\rm 2}  + \xi_{\rm L(R)\sigma}^2  + \Delta_{\rm L(R)}^2},~~~~~\\
&&g^{(0)}_{{\rm FM}\sigma}  (\kms ,\iu\omega_n ) 
=
  - \frac{1}{{\iu\omega_n  - \xi _{{\rm FM}\sigma}  }},\\
&&g^{(0)}_{{\rm FM}\sigma}  (-\kms ,-\iu\omega_n ) 
=
  - \frac{1}{{\iu\omega_n  + \xi _{{\rm FM}\sigma}}},\\
&&\xi _{{\rm L(R)}}
 = \frac{{\klr^{\rm 2} }}{2m} - \mu,  \\ 
&&\xi _{{\rm FM}}^\sigma
= \frac{{\kms^2}}{2m} - \mu  - \sigma \hex.
\eeqa
Note that $\kms$ and $\kmsm$ are independent of each other. 

Finally, we obtain the following formula as, 
\beqa
F
	&=&-\frac{{\left( {mVN_{\rm F} } \right)^2 }}{2}t^4 \Delta ^2 \left( T \right) \nonumber \\
&&\times {\rm Re} 
	\left[\sum_{\iu\omega_n}\left\{
	\frac{\rme^{{\rm sgn}(\omega_n)\iu d
			\sqrt {2m\left( {\mu  + \hex  + \iu\omega _n } \right)}} \rme^{\iu\varphi }}
	{{\omega _n^2  + \Delta ^2 \left( T \right)}}\right.\right.\nonumber\\
&&\left.\left.
 -\frac{\rme^{-{\rm sgn}(\omega_n)\iu d
			\sqrt {2m\left( {\mu  + \hex  + \iu\omega _n } \right)}} \rme^{\iu\varphi }}
	{{\omega _n^2  + \Delta ^2 \left( T \right)}}\right\}\right],
	\label{fcln0}
\eeqa
For $\hex/\mu,~ \omega_n/\mu \ll 1$, eq. (\ref{fcln0}) is approximates as
\beqa
&&F
 \approx
  - \frac{(mV N_{\rm F})^2}{2} 
  t^4 \Delta ^2 (T)\nonumber\\
&&  \times\left(
T \sum_{i\omega_n} 
	\frac{1}{\omega_n^2  + \Delta^2(T)}{\rm e}^{ -2| \omega_n|\df/\vf} 
	\right)\cos\left(\frac{2 \hex \df}{\vf}\right)\cos(\varphi),~~~~~~~\label{fcln}
\eeqa
where $V=S\cdot d$ and $S$ is the cross section between SC and FM. 

\subsection{Disordered system with magnetic scattering}
In the dirty FM with the magnetic scattering, the forth order term is given by
\beqa 
F_{{\rm eff}}
&=& 
	2t^4 \frac{1}{\beta }{\rm Re}[
	\sum_{\{\mbox{\tiny \boldmath $k$}\},\iu\omega_n,\sigma} 
	{\rme^{{\rm  - }\iu \qqbm(\rr - \rl)} }\nonumber\\
	&&
	\times
	f_{\rm R\sigma} (\krs ,\iu\omega_n )
	G_{{\rm FM}\sigma} (\kms ,\iu\omega_n )\nonumber\\
	&&G_{{\rm FM}-\sigma} ( - \kmsm  , - \iu\omega_n)
	f_{\rm L\sigma}^{\rm *} (\kls ,\iu\omega_n )
		\Gamma(\qqbm,\iu\omega_n)],
\eeqa
where $\qqbm=\kms-\kmsm$ in the above notation. 
The Green's function in the dirty FM is  given by
\beqa
G_{{\rm FM}\sigma} (\kbms,\iu\omega _{n} ) 
	&=&
	 \frac{1}{{\iu\omega _{n}  - \xi_{\mbox{\tiny $\kbms$}}  
	  - \Sigma _\sigma  \left( {\kbms,\iu\omega _{n} } \right)}},
\eeqa
where the self-energy, $\Sigma _\sigma  (\kbm,i\omega _{n} ) $, is approximated as,
\beqa
\Sigma _\sigma  (\kbms,\iu\omega _{n} ) 
	&=& - \iu(\frac{1}{2\tau_{\rm imp}}+ \frac{1}{2\taus}){\rm sgn}(\omega _{n} ),~~\\
\taus^{-1}
	&\equiv&2\pi 
	(\frac{J_{\rm H}}{2})^2\frac{V}{N}
N_{\rm F} \left\langle {\delta S^z \delta S^z } \right\rangle.
\eeqa
The diffusive motion of electrons in the FM is described by $\Gamma$ defined as, 
\beqa
\Gamma(\qqbm,\iu\omega_n)\nonumber
&=&
 \sum_m K^m(\qqbm,\iu\omega_n),\\
&\simeq&
 -2N_{\rm F}  \pi \sum_{\mbox{\tiny \boldmath $Q$}} {\frac{{\rm e}^{ - \iu \qqbm(\rr-\rl)} }{{DQ^2  + 2\left| {\omega _{\rm n} } \right| + 2/\taus  \pm \iu 2\hex}}}
 \equiv \Gamma,\label{gamma}\\
K(\qqbm,\iu\omega_n)
&=&\!\!\!\!\!
	\left( { - \frac{{n_i }}{V}\left| u \right|^2  + \left( {\frac{J_{\rm H}}{{2V^{1/2} }}} \right)^2 \left\langle {\delta S^z \delta S^z } \right\rangle } \right)\nonumber\\
	&\times&
	  \sum_{\mbox{\tiny \boldmath $k$}_\sigma} 
	  G_\ua(\mbox{\boldmath $k$}_\ua,\iu\omega_{n} )
	  G_\da(-\mbox{\boldmath $k$}_\da + \qqbm, - \iu\omega_{n}),\\
  &\simeq&
	  \frac{\tauimp - \taus}{\tauimp + \taus }\left(1 - 2\left| {\omega _{n} } \right| \tau - DQ^2 \tau  \mp \iu 2\hex \tau \right).\label{k}
\eeqa
In eqs (\ref{gamma}) and (\ref{k}), we take the dirty limit as, $\wn \tau$, $\hex \tau$, $\vf |Q| \tau$ $\ll$ 1.
Finally, we obtain the free energy in the dirty FM as, 
\beqa
F_{{\rm eff}}
&\simeq&
  - \frac{{t^4 N_{\rm F}^3 \pi ^2 V}}{{4 D\df}}\Delta ^2 (T)\nonumber\\
  &&\times{\mathop{\rm Re}\nolimits} [T\sum\limits_{\iu\omega _{n} } {\frac{1}{{\omega _{n}^{\rm 2}  + \Delta ^2 (T)}}({\rm e}^{ - \alpha ^ +  \df}  + } {\rm e}^{ - \alpha ^ -  \df} ){\rm e}^{\iu\varphi} ],\\
\alpha ^ \pm
	&=&
	\sqrt {\frac{{2(\left| {\omega _{n} } \right| + 1 /\taus  \pm \iu \hex )}}{D}}.
\eeqa

\section{Summation of Matsubara frequency in $\ic$}
A periodic function given by,
\beqa
f(t) 
	&=& -f(t+(2n+1)/T),\nonumber\\
	&=& f(t+2n/T), \label{b1}
\eeqa
can be expanded as, 
\beqa
f(t) 
	&=& T\sum_n \rme^{-\iu \omega_n t},\label{b2}\\
\omega_n
	&=&(2n+1)\pi T.
\eeqa
By taking account of eqs. (\ref{b1}) and (\ref{b2}), an alternating pulse function defined by
\beqa
\delta_{1/T}(t)
	&\equiv& \sum_n (-1)^n\delta(t-n/T), 
\eeqa
is transformed as,
\beqa
\delta_{1/T}(\omega)
	&\equiv& \int\frac{\rmd t}{2\pi}\rme^{\iu \omega t}\delta_{1/T}(t)
	=	\sum_n \delta(\omega-\omega_n).\label{pulse}
\eeqa
Equation (\ref{pulse}) is useful to sum up the Matsubara frequency in eq. (\ref{icln2}) as, 
\beqa
&&T\sum_n \frac{\rme^{-2|\omega_n|\df/\vf}}{\omega_n^2+\Delta^2}\nonumber\\
	&=& \sum_n \int\rmd\omega 
		\frac{\rme^{-2|\omega_n|\df/\vf}}{\omega^2+\Delta^2}\delta(\omega-\omega_n)\\
	&=& \int\frac{\rmd t}{2\pi} \delta_{1/T}(t) 
				\int\rmd\omega 
					\frac{\rme^{- 2|\omega|\df/\vf + \iu \omega t}}{\omega^2+\Delta^2},\\
	&=& \int\frac{\rmd t}{2\pi} \delta_{1/T}(t) 
				\int_0^\infty\rmd \omega 
					\frac{\rme^{- 2\omega\df/\vf + \iu \omega t}
							 +\rme^{- 2\omega\df/\vf - \iu \omega t}}{\omega^2+\Delta^2},\\
	&=& \frac{1}{\Delta} \int\frac{\rmd t}{2\pi} \delta_{1/T}(t) 
				\int_0^\infty\rmd x \sin x \nonumber\\
				&&\times
					\left[
						\frac{1}{x+(2d/\vf-\iu t)\Delta}+\frac{1}{x+(2d/\vf+\iu t)\Delta}
					\right],\\
	&=& \frac{1}{\Delta} \int\frac{\rmd t}{2\pi} \delta_{1/T}(t) 
				\int_0^\infty\rmd x \sin x \nonumber\\
				&&\times
					\left[
						\frac{x+ d/\xi_0}{(x+ d/\xi_0)^2+(t\Delta)^2}
					\right],\\
	&=& \frac{1}{2\pi \Delta}
				\int_0^\infty\rmd x \sin x \nonumber\\
				&&\times\sum_n(-1)^n
					\left[
						\frac{x+ d/\xi_0}{(x+ d/\xi_0)^2+n^2(\Delta/T)^2}
					\right],\\
	&=& \frac{1}{2\pi\Delta}
			\int_0^{\infty}\rmd x ~\sin x 
				\left[
				\frac{\pi T}{\Delta}{\rm cosech}\left(\frac{\pi T x}{\Delta}
       +\frac{\df}{\xit}\right)
				\right],\label{b22}~~~~~~
\eeqa
where $\xit\equiv \vf/2\pi T$ and $\xi_0\equiv \vf/2\Delta$.
The following relation is used to obtain the forth equality in the above equation,
\beqa
&&\int_0^{\infty}\rmd\omega ~\frac{\rme^{-p\omega}}{\omega^2+1}
	=\!\!\! \int_0^{\infty}\rmd x ~\frac{\sin x}{x+p}\nonumber\\
	&=&\!\!\! -\cos(p){\rm si}(p) +\sin(p){\rm ci}(p).
\eeqa

In the limit of $\df/\xit\gg 1$, eq. (\ref{b22}) is approximated as,
\beqa
&&\frac{1}{2\pi\Delta}
			\int_0^{\infty}\rmd x ~\sin x 
				\left[
				\frac{\pi T}{\Delta}{\rm cosech}\left(\frac{\pi T x}{\Delta}
       +\frac{\df}{\xit}\right)
				\right],\\
&\simeq& \frac{1}{2\pi\Delta}
			\int_0^{\infty}\rmd x ~\sin x \times
				\left[
				\frac{2\pi T}{\Delta}\exp\left(-\frac{\pi T x}{\Delta}-\frac{\df}{\xit}\right)
				\right],~~~\nonumber\\
&&\propto \exp\left(-\frac{\df}{\xit}\right)~~~\mbox{for}~\df/\xit\gg 1.
\eeqa

On the other hand, in the limit of $T\rightarrow 0$ K,
\beq
\lim_{T\rightarrow 0 {\rm K}}{\rm cosech}
\left[\frac{\pi T}{\Delta}\left(x+\frac{2\df\Delta}{\vf }\right)\right]
	= \frac{\Delta}{\pi T}\frac{1}{x+\df/\xi_0}, 
\eeq
where $\xi_0\equiv \vf/2\Delta$.
Then, 
\beqa
&& \lim_{T\rightarrow 0}\kbt \sum_{\omega_n}
        \frac{\rme^{-2|\omega_n|\df/\vf}}{\omega ^{2}_{n}+\Delta ^{2}},\\
&=& \frac{1}{2\pi\Delta}
			\int_0^{\infty}\rmd x ~	\frac{\sin x}{x+ \df/\xi_0}\nonumber\\
&=& \frac{1}{2\pi\Delta}
			\left[ -\cos\left(\frac{\df}{\xi_0}\right){\rm si}\left(\frac{\df}{\xi_0}\right)
			+\sin\left(\frac{\df}{\xi_0}\right){\rm ci}\left(\frac{\df}{\xi_0}\right)\right],\\
	&\propto& \frac{1}{\df/\xi_0}. 
\eeqa
}

\begin{figure}[p]
\includegraphics[height=8cm]{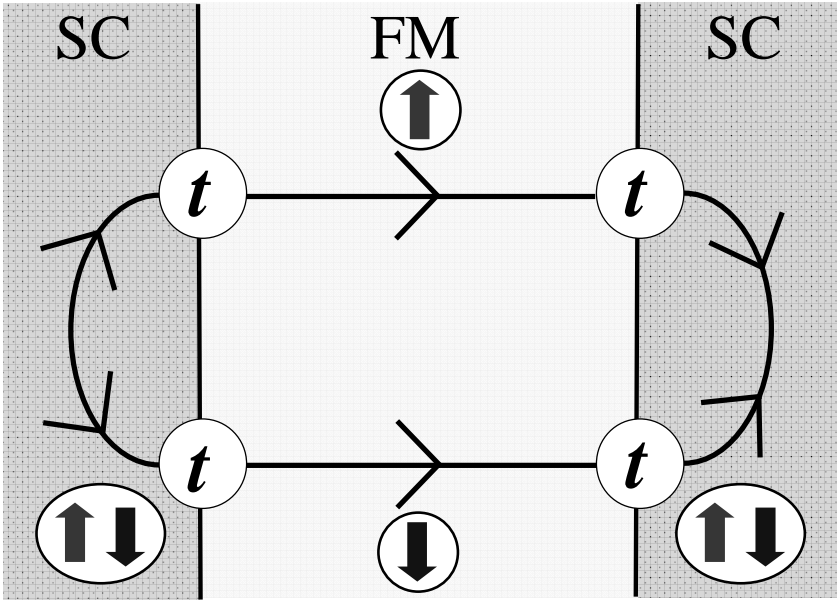}
\caption{Schematic figure of SFS junction and the fourth order diagram for $F$. The circles with up- and down arrows in SC indicate a Cooper pair, and the circle with up (down) arrow in FM indicate the up-spin electron (down-spin electron). A tunneling matrix element is denoted by $t$.}
\label{4th}
\end{figure}
\begin{figure}[p]
\includegraphics[height=8cm]{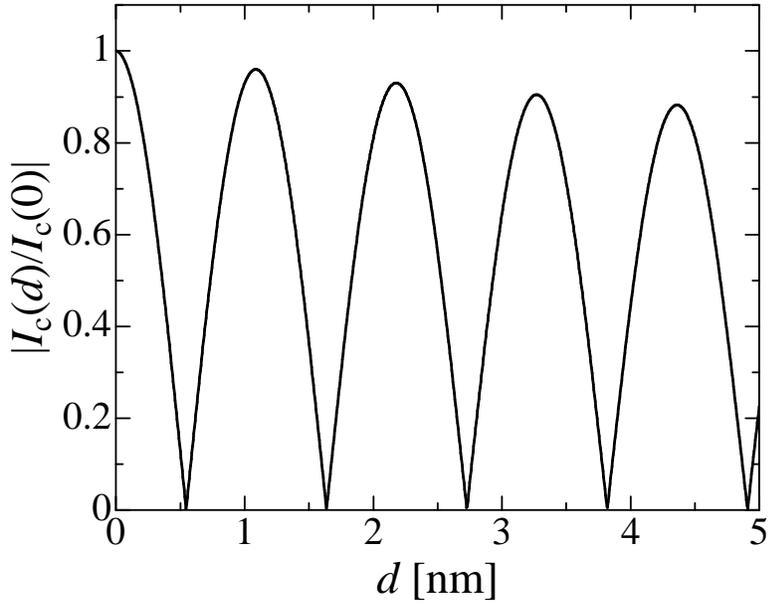}
\caption{$\ic$ in the clean system for $\vf$=2.5$\times$10$^{5}$ m/s, $h_0$=0.36 eV, $\Delta$=1.5 meV, $T$=4 K, and $\tfm/\tc$=10.}
\label{ddepcln}
\end{figure}
\begin{figure}[p]
\includegraphics[height=8cm]{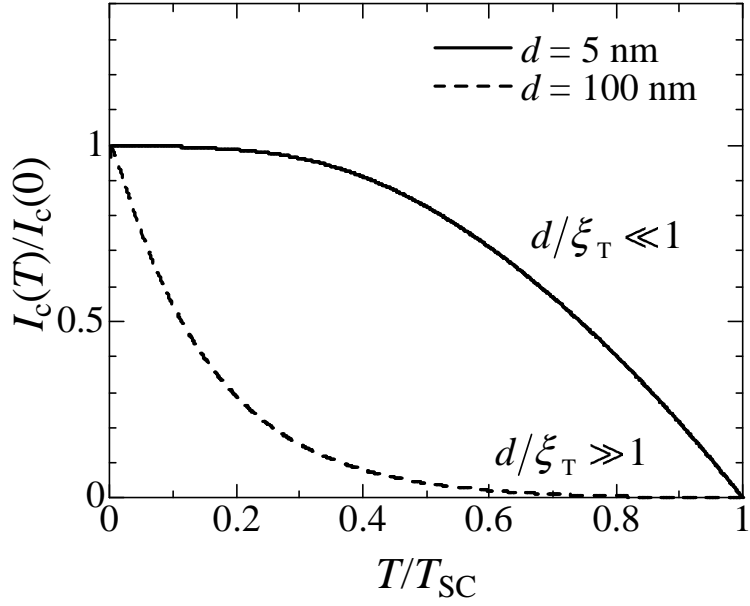}
\caption{
The critical current $\ic$ as a function of $T$ is plotted for $\vf$=2.5$\times$10$^{5}$ m/s and $\tc$=10 K with $d$=5 nm by the solid line and with $d$=100 nm by the broken line. Here, $\xit$ is greater than $\vf/2\pi\tc\simeq$30 nm. Hence, the solid line always satisfies $\df$/$\xit<$1 and becomes like the $\ic$-$T$ curve in the SIS junction, while the broken line seems to be that in the SNS junction due to $\df$/$\xit>$1 for $T/\tc\gsim0.3$. 
}
\label{tdep}
\end{figure}
\begin{figure}[p]
\includegraphics[width=8cm]{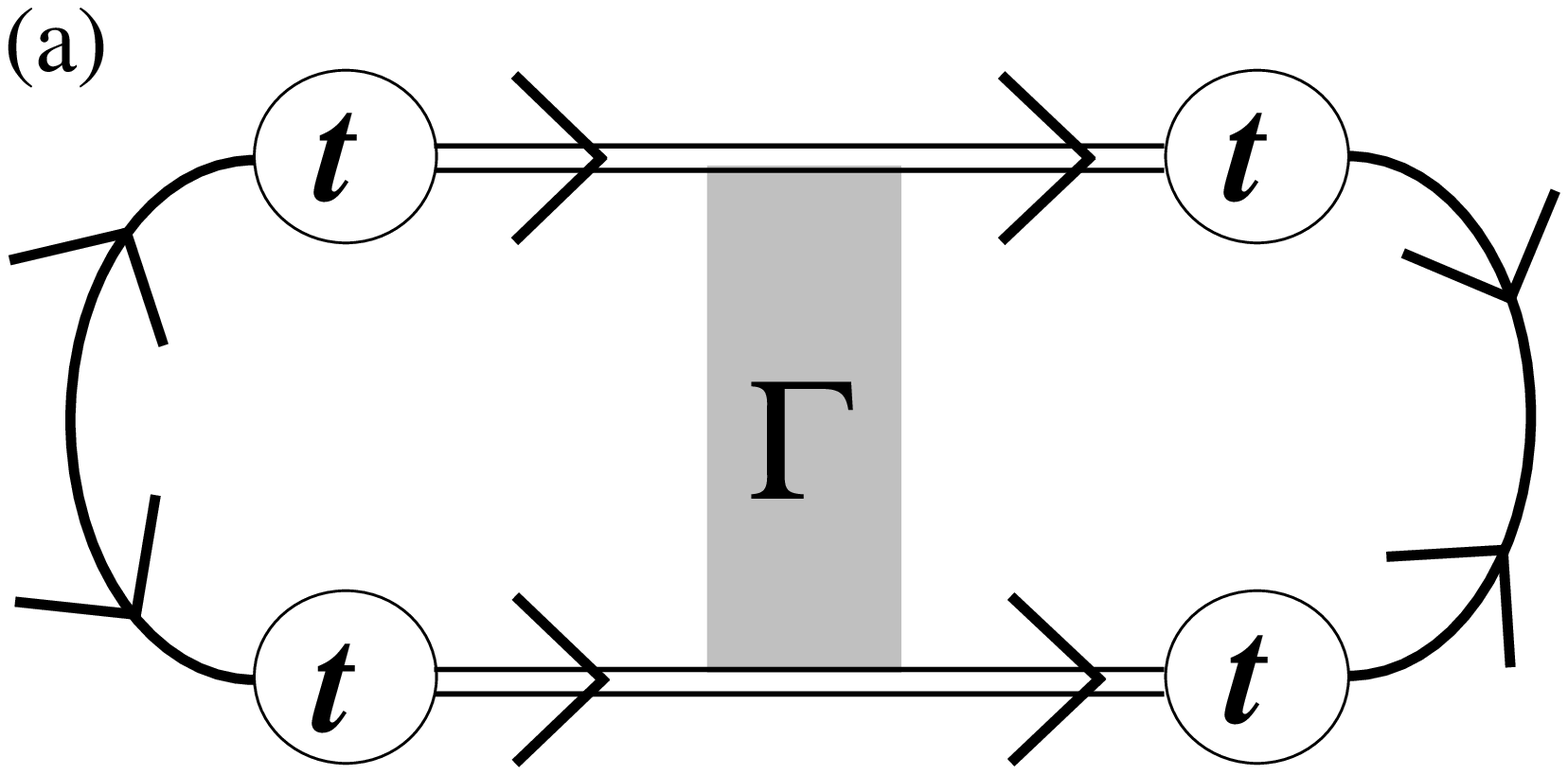}

\vspace{5mm}
\includegraphics[width=10cm]{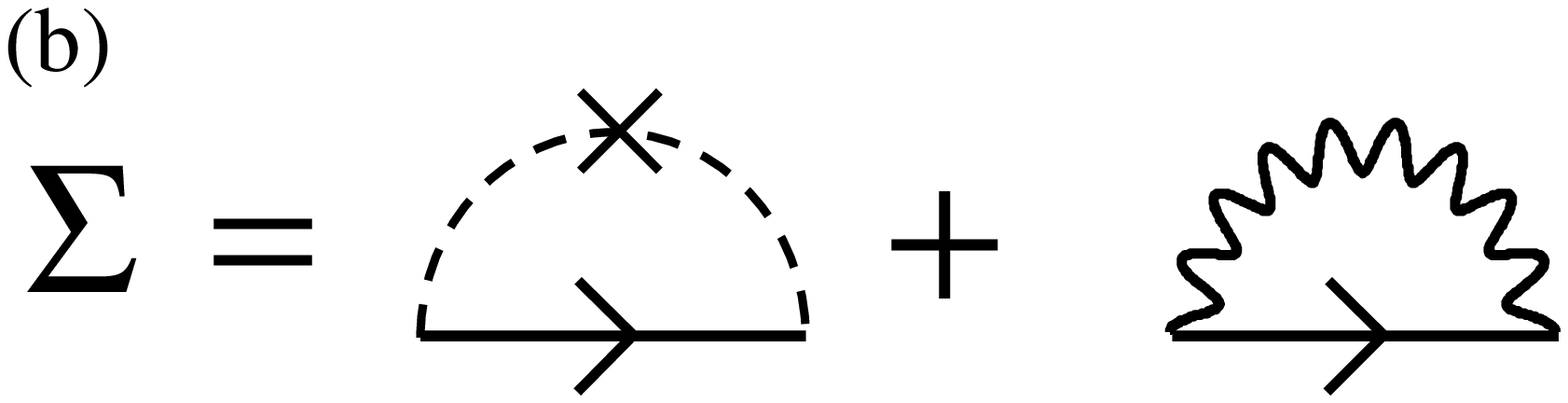}
\caption{
(a) The fourth order diagram contributing to the Josephson current in the dirty FM. $t$ is a tunneling matrix element. The double solid lines represent the Green function with the self-energy, $\Sigma \left( {\omega _{\rm n} } \right) $, in the Born approximation. 
(b) The self-energy by non-magnetic and magnetic scatterings in the Born approximation.
}
\label{diffuse}
\end{figure}
\begin{figure}[p]
\begin{center}
\includegraphics[height=1.5cm]{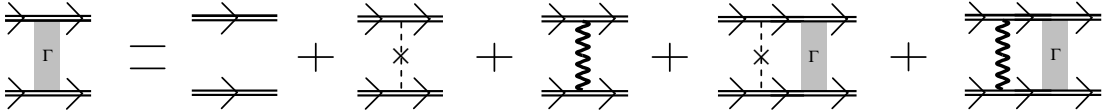}
\caption{
Diagram of diffusive electrons in the FM by non-magnetic and magnetic scatterings.
}
\end{center}
\label{ladder}
\end{figure}
\begin{figure}[p]
\begin{center}
\includegraphics[width=9cm]{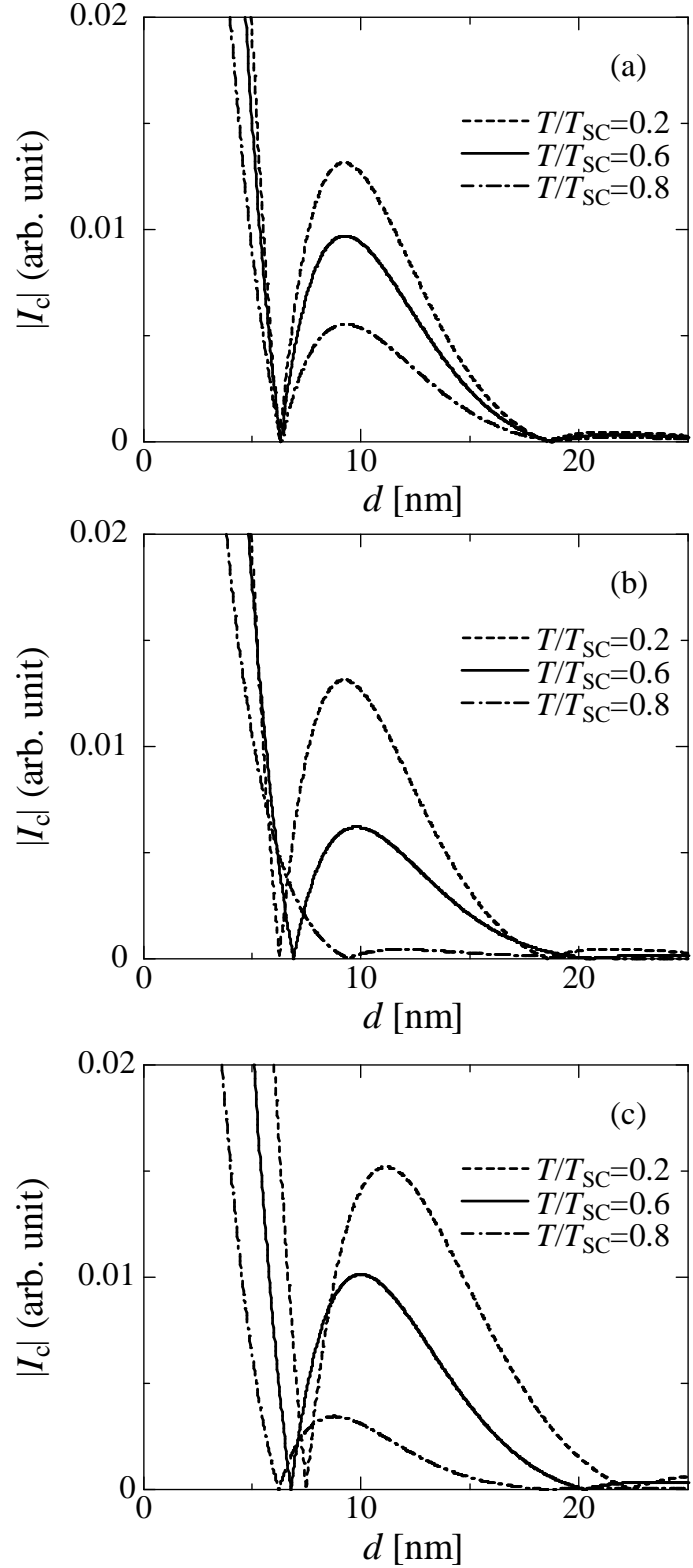}
\caption{The absolute values of Josephson critical current as a function of $d$ for $\vf$=2.5$\times10^{5}$ m/s, and 
(a) $T_{\rm FM}/T_{\rm SC}$=10, $h_{0}$=30 meV, $\tauimp$=2$\times10^{-14}$ s, $\tau_{s0}$=$10^{-12}$ s, 
(b) $T_{\rm FM}/T_{\rm SC}$=1,  $h_{0}$=30 meV, $\tauimp$=2$\times10^{-14}$ s, $\tau_{s0}$=$10^{-13}$ s, 
(c) $T_{\rm FM}/T_{\rm SC}$=1,  $h_{0}$=100  meV, $\tauimp$=2$\times10^{-13}$ s, $\tau_{s0}$=$10^{-13}$ s. 
The ferromagnetic and superconducting transition temperatures are denoted by 
$T_{\rm FM}$ and $T_{\rm SC}$, respectively. 
It is interesting that the period increases with $T$ in Fig. \ref{ddep} (b), while that decreases with $T$ in Fig. \ref{ddep} (c).}
\label{ddep}
\end{center}
\end{figure}
\begin{figure}[p]
\begin{center}
\includegraphics[width=10cm]{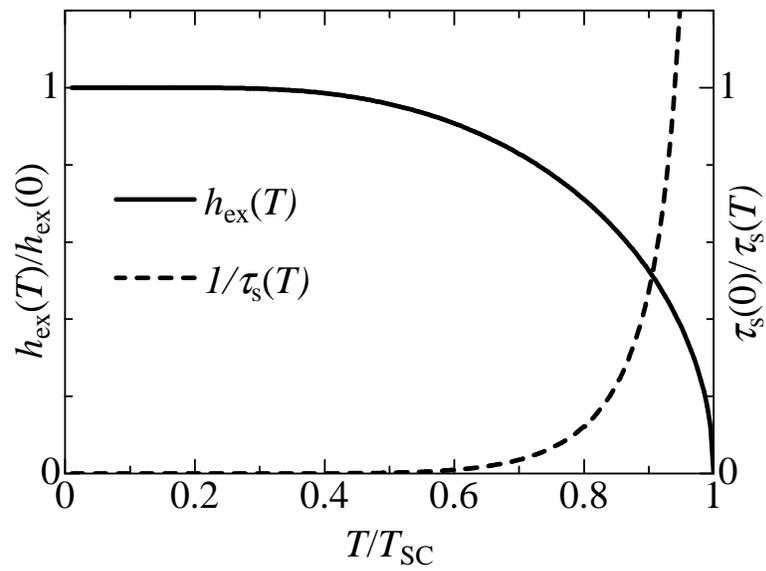}
\caption{The temperature dependence of $\taus$ and $\hex$ for $\tfm/\tc=1$ are plotted  by the broken and solid lines, respectively. }
\label{taus}
\end{center}
\end{figure}
\begin{figure}[p]
\begin{center}
\includegraphics[width=10cm]{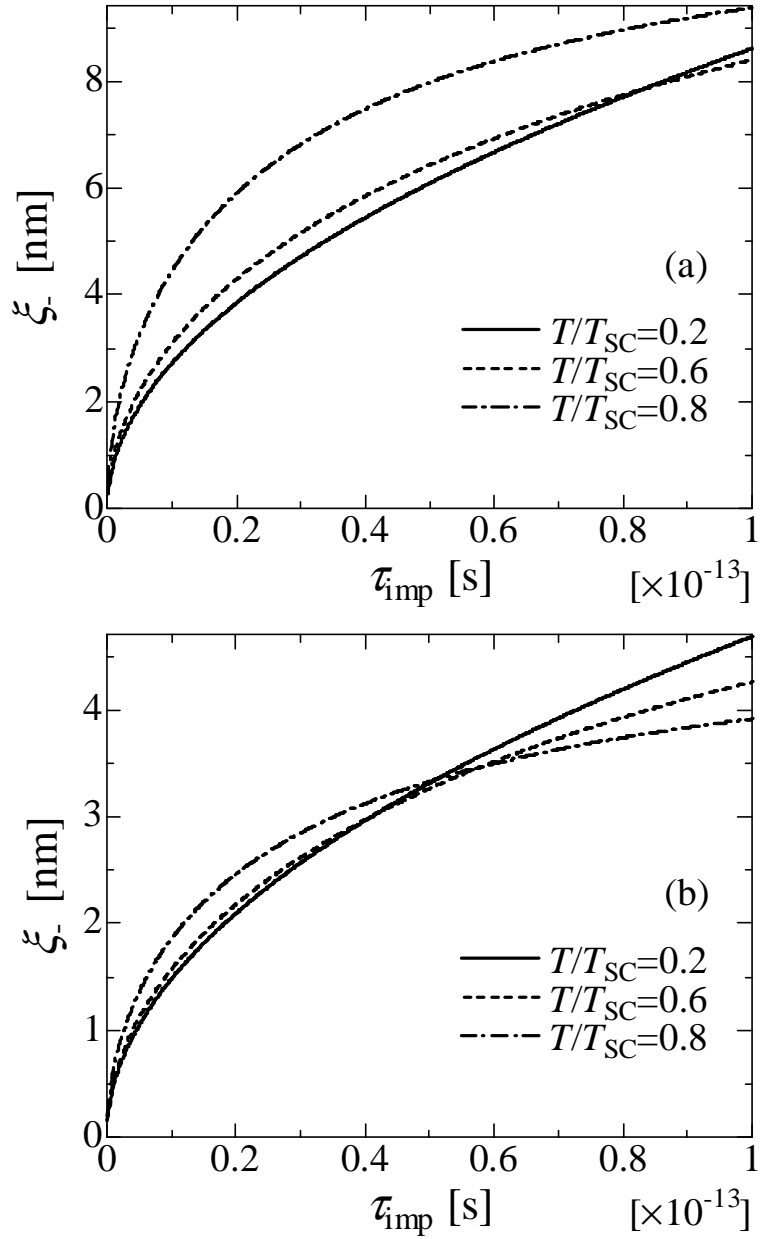}
\caption{The period of oscillation as a function of $\tauimp$ with $\vf$=2.5$\times10^{5}$ m/s, $T_{\rm FM}/T_{\rm SC}$ =1, and $\tau_{s0}$=10$^{-13}$ s, for (a) $h_{0}$=30 meV, $\tauimp$=2$\times10^{-14}$ s, (b) $h_{0}$=100 meV, $\tauimp$=2$\times10^{-13}$ s.}
\label{xi}
\end{center}
\end{figure}
\begin{figure}[p]
\begin{center}
\includegraphics[width=10cm]{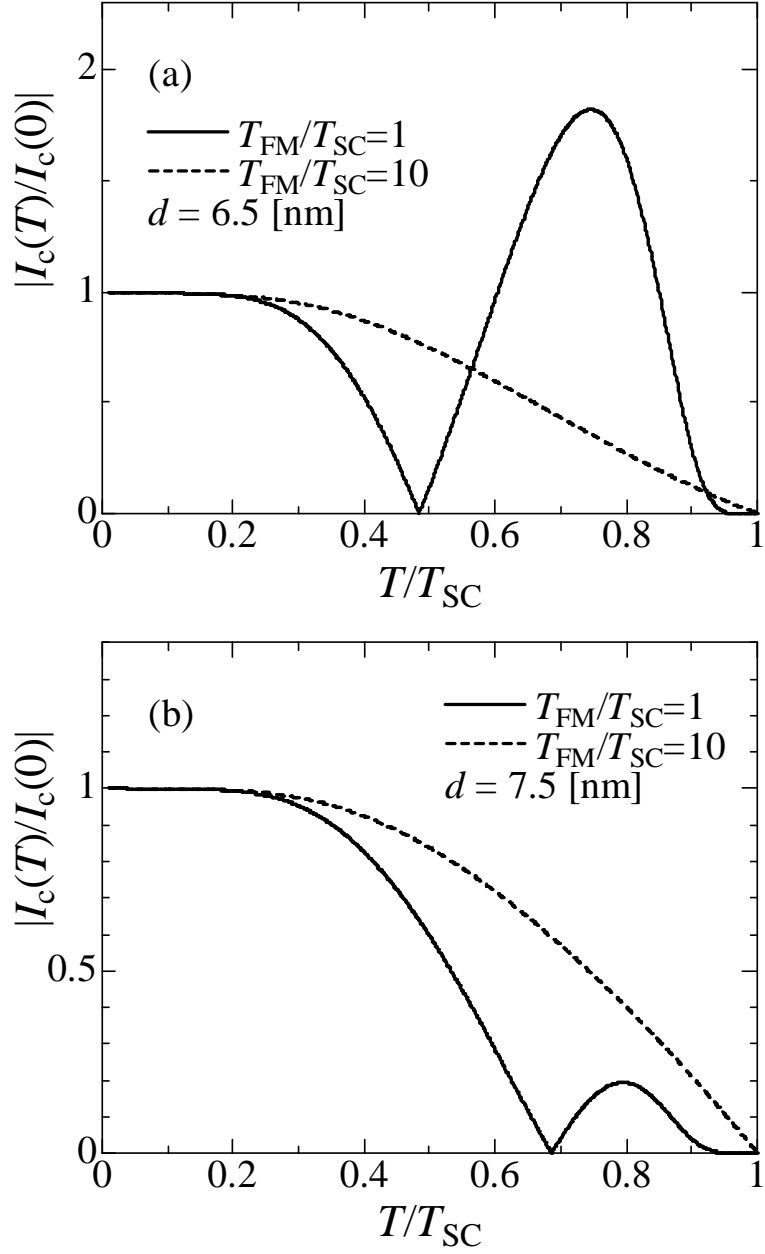}
\caption{Normalized Josephson critical current as a function of $T/T_{\rm SC}$ in the case of (a) $d$=6.5 nm and (b) $d$=7.5 nm, for $T_{\rm FM}/T_{\rm SC}$=1, $h_{0}$=30 meV, $\tauimp$=2$\times10^{-14}$ s, $\tau_{s0}=10^{-13}$ s.}
\label{tdrt}
\end{center}
\end{figure}
\begin{figure}[p]
\begin{center}
\includegraphics[width=10cm]{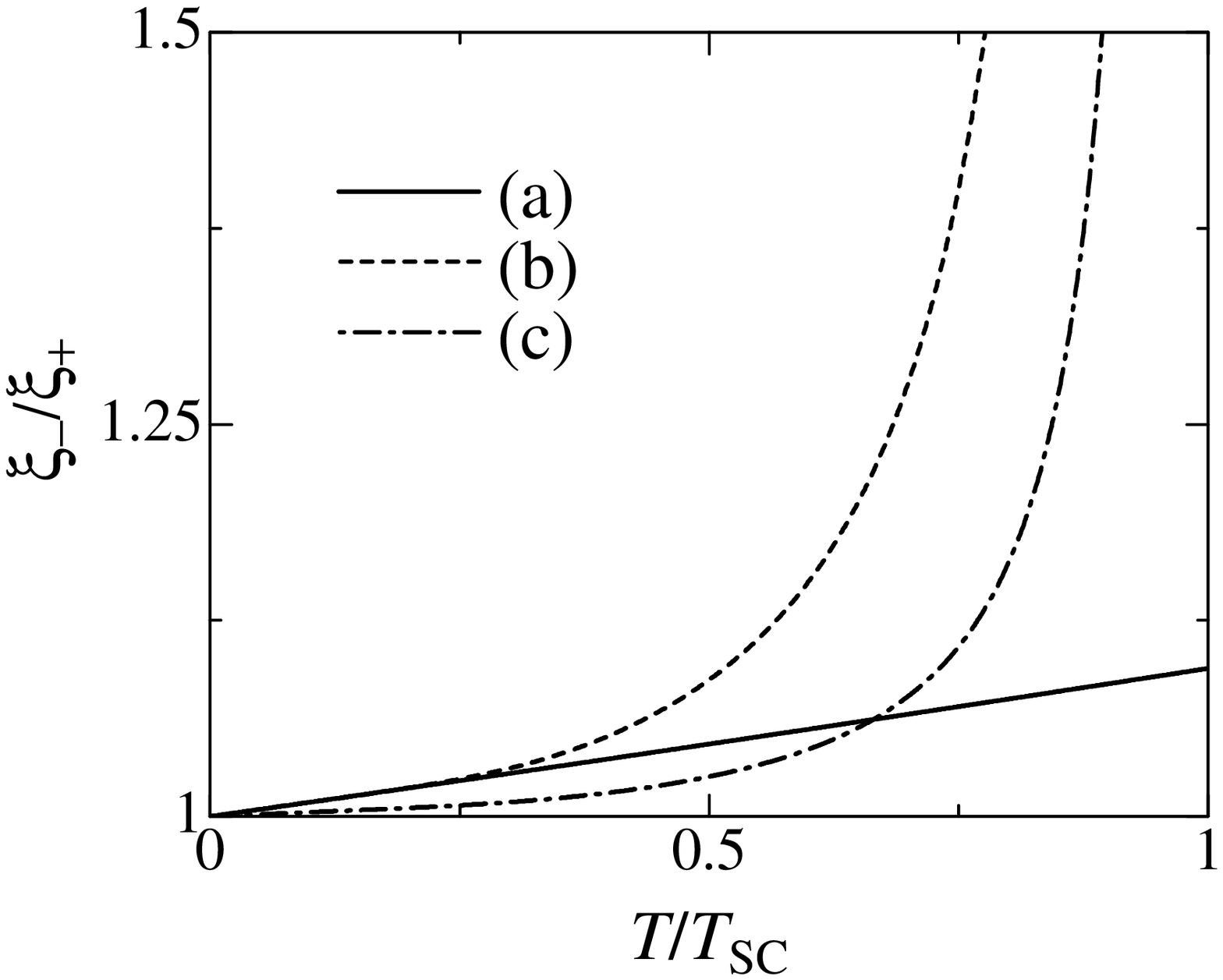}
\caption{The ratios of the period and the decay length, $\xi_-/\xi_+$, are plotted; 
 (a) by the solid line for 
$T_{\rm FM}/T_{\rm SC}$=10, $h_{0}$=30 meV, $\tauimp$=2$\times10^{-14}$ s, $\tau_{s0}$=$10^{-12}$ s, 
 (b) by the broken line for 
$T_{\rm FM}/T_{\rm SC}$=1,  $h_{0}$=30 meV, $\tauimp$=2$\times10^{-14}$ s, $\tau_{s0}$=$10^{-13}$ s, 
and (c) by the dotted broken line for $T_{\rm FM}/T_{\rm SC}$=1,  $h_{0}$=100  meV, $\tauimp$=2$\times10^{-13}$ s, $\tau_{s0}$=$10^{-13}$ s. 
The other parameters are set to $\omega_{n=0}$ and $\vf$=2.5$\times10^{5}$ m/s.
The ratio is dominated by $T$ in the case of (a) or around $T\simeq$ 0 K in the cases (b) and (c), since $\pi T$ is larger than $1/\taus$. 
On the other hand, below $\tc$ in the cases of (b) and (c), the ratio exponentially increases due to $T$-dependence of $1/\taus$, which is much larger than $T$. 
}
\label{ratio}
\end{center}
\end{figure}
\end{document}